\documentclass[12pt,draftcls,onecolumn,journal]{IEEEtran}
\usepackage{graphicx,times,cite,amsmath, amssymb, epsfig, array}
\usepackage{multirow}

\usepackage{array}
\usepackage{textcomp}
\newlength{\figwidth}
\begin{document}
\title{\LARGE Opportunistic Cooperative Channel Access in Distributed Wireless Networks with Decode-and-Forward Relays}
\author{Zhou~Zhang, Shuai Zhou, and Hai Jiang\\Dept. of Electrical \& Computer Engineering, University of Alberta, Canada
\vspace{-6mm}}

\maketitle
\begin{abstract}
This letter studies distributed opportunistic channel access in a wireless network with decode-and-forward relays. All the sources use channel contention to get transmission opportunity. If a source wins the contention, the channel state information in the first-hop channel (from the source to its relay) is estimated, and a decision is made for the winner source to either give up the transmission opportunity and let all sources start a new contention, or transmit to the relay. Once the relay gets the traffic, it may have a sequence of probings of the second-hop channel (from the relay to the destination). After each probing, if the second-hop channel is good enough, the relay transmits to the destination and completes the transmission process of the source; otherwise, the relay decides either to give up and let all sources start a new contention, or to continue to probe the second-hop channel. The optimal decision strategies for the two hops are derived in this letter. The first-hop strategy is a pure-threshold strategy, i.e., when the first-hop channel signal-to-noise ratio (SNR) is more than a threshold, the winner source should transmit to the relay, and subsequently the second-hop strategy should let the relay keep probing the second-hop channel until a good enough second-hop channel is observed. Simulation results show that our scheme is beneficial when the second-hop channels have larger average SNR.
\end{abstract}
\begin{keywords}
Decode-and-forward relaying, opportunistic transmission, optimal stopping, throughput maximization.
\end{keywords}

\section{Introduction}

In a distributed wireless network such as ad hoc network, normally the traffic sources use channel contention to obtain transmission opportunity. For example, if a source has traffic to send, it sends a request-to-send (RTS) to its destination, and if the RTS is successfully received, the destination sends back a clear-to-send (CTS), and then the source can transmit its traffic, even if its channel to the destination is poor. To efficiently utilize the wireless channel, it may be better if a source could give up its transmission opportunity when its channel is not good, i.e., it does not transmit upon reception of CTS, and thus all sources immediately start a new round of channel contention. The challenge is: when should a channel be called ``good channel"? This challenge can be addressed easily for a centralized network coordinated by a central controller. For example, in a cellular network, the base station (BS) is the central controller, and collects channel state information of all users for each channel. Then the BS can pick up the best user, which has the best channel gain, to utilize a channel, referred to as {\it opportunistic channel access} or {\it channel-aware scheduling}. However, in a distributed network, no such central controller exists to decide which user has the best channel gain. Rather, each user needs to decide based only on its local observation (i.e., its own channel gain) and channel statistics of other users' channels. The challenge of opportunistic channel access in a distributed network was addressed in \cite{zheng_dong}. A source first sends a probe (e.g., RTS) to its destination for channel contention. If the contention is successful, the destination estimates the channel signal-to-noise ratio (SNR) and feedbacks (e.g., by using CTS) to the source. If the channel SNR is less than a threshold value, which can be numerically calculated based on the statistics of users' channels, then the source gives up its transmission opportunity; otherwise, the source transmits its traffic using the maximal achievable transmission rate of the probed channel. As follow-ups of \cite{zheng_dong}, the works in \cite{imperfect_information}, \cite{Ge_Zheng_2009}, \cite{Tan_Zheng_2010}, \cite{two_level} investigate opportunistic channel access in a distributed wireless network when channel information is imperfect, when multiple transmissions can be successfully received, when there exists a delay constraint, and when a fine channel estimation could be available, respectively. For wireless relaying networks, distributed opportunistic channel access is investigated in \cite{gong_xiaowen} and \cite{zhangzhou} with decode-and-forward (DF) relays and  amplify-and-forward (AF) relays, respectively. Next, the work in \cite{gong_xiaowen} is introduced since this letter is a follow-up of the work in \cite{gong_xiaowen}. A distributed DF relaying network is considered in \cite{gong_xiaowen}, in which each source-destination pair is aided by a DF relay. If a source has traffic to send, it sends a probing packet, and based on reception of the probing packet, the information of channel SNR in the first hop (from the source to its relay and to the destination) is obtained. Then the source decides to 1) give up, or 2) transmit with direct link, or 3) continue to probe the second hop (from the relay to the destination). If the source decides to probe the second hop, then the channel SNR of the second hop  is estimated, and it is decided either to give up or to transmit (by using direct link or relay link, whichever has better utility).

In \cite{gong_xiaowen}, when the source decides to proceed to probe the second hop, if the second-hop channel SNR is not good, it is likely that the source will give up the transmission opportunity and all sources start a new contention. However, since the second-hop channel is time-varying, a natural question is: if a second-hop channel is poor, is it beneficial to let the relay wait until the second-hop channel becomes better? The rationale for this question is: if the source gives up when the second-hop channel is poor, then it still takes time for the system to have a successful channel contender later, and the successful channel contender may not have good channel SNRs in two hops. So it is possible to have benefits by using a relay-waiting scheme, i.e., letting the relay wait until the second-hop channel becomes better. This letter targets at an answer to the aforementioned question. In specific, we first derive an optimal strategy for the relay-waiting scheme. Then performance of the relay-waiting scheme and the scheme in \cite{gong_xiaowen} are compared by simulations, which demonstrate that the relay-waiting scheme can be beneficial when the second-hop channels have larger average SNR.


\section{System Model}\label{SystemModel}

Consider a distributed DF relay network that includes a number, $M$, of source-destination pairs. Each source-destination pair has a relay assigned. First consider the case with direct links from sources to destinations. Similar to \cite{gong_xiaowen}, to probe the first-hop channels, a source can send a probing packet, and if there is no collision, the probing packet is received by both its relay and its destination. The relay and the destination then can estimate the channel SNRs from the source to themselves. Then the relay reports its channel SNR information to the destination, and the destination makes decision for the first hop (give up or transmit). For this case, by reception of the reporting message from the relay, the destination can estimate the channel SNR from the relay to itself, and thus, the destination has complete channel SNR information for the two hops: from the source to the relay, from the source to itself, and from the relay to itself. Then the destination can calculate the achievable end-to-end transmission rate given as $R$ between the source and itself. Therefore, although the communication from the source to the destination is with two hops, it can be treated as a virtual one-hop communication with achievable rate $R$. So the same method as that in \cite{zheng_dong} (which deals with single-hop ad hoc networks) can be used to find out the opportunistic channel access strategy. Therefore, in this letter, we investigate the case without direct links between each source-destination pair. Assume channels in the first hop (i.e., from sources to their relays) follow independent and identically distributed (i.i.d.) Rayleigh fading with average received SNR being $\rho_f$, while the channels in the second hop (i.e., from relays to destinations) follow i.i.d. Rayleigh fading with average received SNR being $\rho_g$.

The $M$ sources use a channel contention procedure similar to those in \cite{zheng_dong}\cite{gong_xiaowen}\cite{zhangzhou}, as follows. At a minislot (the duration of which is denoted as $\sigma$), each source sends an RTS with probability $p$ to its relay. So at each minislot, if no source transmits, i.e., the minislot is idle (the probability is $(1-p)^M$), then all sources start a new channel contention in next minislot; if more than one source send RTS (the probability is $1-(1-p)^M - Mp(1-p)^{M-1}$), it means that transmissions of the sources collide with each other, and thus, all the sources start a new channel contention after a time-out value (the duration of which is equal to CTS duration) following the collision; if only one source sends RTS (with probability $Mp(1-p)^{M-1}$), then we call the source {\it a winner source}. Define {\it an observation} as the interval from the starting point of the channel contention until a winner source appears (i.e., its RTS is successfully received by its relay). 
The average duration of an observation can be calculated as $\tau_1 = \frac{(1-p)^M}{Mp(1-p)^{M-1}}\cdot \sigma+\frac{1-(1-p)^M-Mp(1-p)^{M-1}}{Mp(1-p)^{M-1}}(\tau_{RTS}+\tau_{timeout}) + \tau_{RTS}$, in which $\tau_{RTS}$ and $\tau_{timeout}$ are RTS and timeout durations, respectively.

At the end of an observation (say, observation $n$), the winner source's relay can estimate the channel SNR from the winner source to itself by the RTS reception, and it decides from two options: 1) option {\it give-up}: to give up the transmission opportunity, and notify the source of the decision by sending back a CTS. This CTS is also received by other sources. Thus, subsequently all sources can start a new contention. 2) option {\it stop}: to {\it stop} the process and utilize the transmission opportunity, and send back a CTS to notify the decision. In the CTS, a transmission rate denoted $R_n$ is also indicated for transmission from the winner source to the relay. Then the winner source transmits for duration of a channel coherence time denoted as $\tau_d$ by using transmission rate $R_n$. The optimal value of $R_n$ is derived in Section \ref{SchemefortheFirstHop}.

For observation $n$, if the winner source stops, denote reward $Y_n$ as the total amount of traffic that is sent by the winner source and received by its destination, and denote $T_n$ as the time duration from observation $1$ until observation $n$ plus the time used for transmissions in the two hops.
Denote $N$ as the {\it stopping time}, i.e., the winner sources in the first $N-1$ observations do not stop, and the winner source in the $N$th observation stops. This letter targets at an optimal stopping time denoted as $N^*$, which makes the system achieve the maximal system throughput, i.e.,
\begin{equation}\label{e:original}
N^* =\arg  \sup\limits_{N\ge 0}\frac{\mathbb{E}[Y_N]}{\mathbb{E}[T_N]}
\end{equation}
where $\mathbb{E}[\cdot]$ means expectation. $N^*$ is also referred to as {\it optimal stopping strategy}. Based on \cite[Chapter 6]{ferguson2006optimal}, we can transform problem (\ref{e:original}) into a problem that maximizes reward $ Y_N\!-\!\lambda T_N$ with $\lambda>0$. In specific, for $\lambda>0$, an optimal strategy denoted $N^*(\lambda)$ should be found, which maximizes expected reward of the transformed problem:
\begin{equation}\label{e:transform}
U(\lambda)=\sup\limits_{N(\lambda)\ge0}\{\mathbb{E}[Y_{N(\lambda)}]-\lambda \mathbb{E}[T_{N(\lambda)}]\}.
\end{equation}
Then if we find a $\lambda^*$ such that $U(\lambda^*)=0$, then an optimal strategy of problem (\ref{e:original}) is in the form of $N^*(\lambda)$ with $\lambda = \lambda^*$ \cite{ferguson2006optimal}.

Next we find optimal strategy for problem (\ref{e:transform}), which includes two parts: the optimal second-hop strategy and optimal first-hop strategy, discussed in the subsequent two sections.

\section{Strategy for the Second Hop}\label{SchemefortheSecondHop}

Consider observation $n$. Here we first try to find the optimal strategy for the second hop, i.e., we assume the winner source stops and transmits to its relay with rate $R_n$. For the second hop, the relay should find out its best strategy. The relay first sends an RTS to the destination, and the destination estimates the second-hop channel SNR denoted $r_g$ and feedbacks a CTS that includes the channel SNR information, referred to as a {\it channel probing}. If the achievable second-hop transmission rate, given as $\log_2(1+r_g)$, is not less than $R_n$, then the relay transmits to the destination by using transmission rate $R_n$ with duration $\tau_d$; otherwise, the relay may decide to give up or to continue channel probing. If the relay decides to give up, all sources start a new channel contention. If the relay decides to continue channel probing, then the relay waits for channel coherence time $\tau_d$ and has a new RTS-CTS exchange with the destination (a new channel probing), and transmits if the achievable second-hop transmission rate is not less than $R_n$, or decides to give up or to continue channel probing otherwise. This procedure is repeated until the relay either transmits or gives up. It can be seen that there are a sequence of decisions in the second hop, which makes the optimal second-hop strategy challenging. To address the challenge, we review second-hop strategies from a new perspective, as follows.

Denote $S_l$ as the second-hop strategy that the relay can have up to $l$ channel probings of its channel to the destination. So if the relay cannot find a second-hop channel realization with achievable rate not less than $R_n$ within $l$ channel probings, the relay is forced to give up. Denote $V^l(\lambda)$ (which is a function of $\lambda$) as the net reward of strategy $S_l$. Therefore, the optimal second-hop strategy should achieve net reward $\max \{ {\mathbb E}[V^1(\lambda)], {\mathbb E}[V^2(\lambda)],...,{\mathbb E}[V^\infty(\lambda)]\}$.

The net reward expectation of strategy $S_1$ is
 \begin{equation}\label{e:s1_reward}
  \begin{split}
    \mathbb{E}[V^1(\lambda)]&=\text{Pr}[r_g^1 \geq r_n](R_n\tau_d-\lambda\tau_2)  + \text{Pr}[r_g^1 < r_n](-\lambda(\tau_{RTS} + \tau_{CTS})) \\
        &=(1-F_g(r_n))(R_n\tau_d-\lambda\tau_2) + F_g(r_n)(-\lambda(\tau_{RTS} + \tau_{CTS}))
  \end{split}
\end{equation}
where $\text{Pr}[\cdot]$ means probability of an event, $\tau_{CTS}$ is CTS transmission duration, $\tau_2 = \tau_{RTS}+\tau_{CTS}+\tau_d$ is the time cost for probing and transmission in the second hop, $F_g(\cdot)$ is the cumulative distribution function (CDF) of the second-hop channel SNR (the subscript $g$ stands for the second hop), $r_g^1$ is the second-hop channel SNR in the first channel probing, $r_n \triangleq 2^{R_n}-1$ is the minimum required SNR of the second hop for achievable transmission rate $R_n$.

The net reward expectation of strategy $S_{\infty}$ is
\begin{equation}\label{e:sinfty_reward}
  \begin{split}
    &\mathbb{E}[V^{\infty}(\lambda)] \\
    =&\text{Pr}[r_g^1 \geq r_n](R_n\tau_d-\lambda\tau_2) + \text{Pr}[r_g^1 < r_n](\mathbb{E}[V^{\infty}(\lambda)]-\lambda\tau_2) \\
    =& (1-F_g(r_n))(R_n\tau_d-\lambda\tau_2)+ F_g(r_n)(\mathbb{E}[V^{\infty}(\lambda)]-\lambda\tau_2).
  \end{split}
\end{equation}
From (\ref{e:s1_reward}) and (\ref{e:sinfty_reward}), we have
\begin{equation}
    \mathbb{E}[V^{\infty}(\lambda)]-\mathbb{E}[V^1(\lambda)]  
  =   F_g(r_n)(\mathbb{E}[V^{\infty}(\lambda)]- \lambda\tau_d).
  \label{force_to_stop_or_not}
\end{equation}

\subsection{Case with $\mathbb{E}[V^{\infty}(\lambda)] \ge \lambda\tau_d $}\label{s:vinfmore}

If $\mathbb{E}[V^{\infty}(\lambda)]\ge \lambda\tau_d$ , then $\mathbb{E}[V^{\infty}(\lambda)] \geq \mathbb{E}[V^1(\lambda)]$. Now we compare $\mathbb{E}[V^{\infty}(\lambda)]$ with $\mathbb{E}[V^l(\lambda)]$, $l\ge 1$.

We have
\begin{multline}
{\mathbb E}[V^l(\lambda)]=\text{Pr}[r_g^1\ge r_n](R_n\tau_d-\lambda \tau_2)
+\text{Pr}[r_g^1< r_n, r_g^2\ge r_n](R_n\tau_d-2\lambda \tau_2) + ...\\
+\text{Pr}[r_g^1<r_n,...,r_g^{l-1}<r_n,r_g^l\ge r_n](R_n\tau_d-l\lambda \tau_2)\\
+\text{Pr}[r_g^1<r_n,...,r_g^{l-1}<r_n,r_g^l< r_n] (-(l-1)\lambda \tau_2-\lambda(\tau_{RTS}+\tau_{CTS}))
\end{multline}
in which $r_g^1,r_g^2,...,r_g^l$ are channel SNRs of 1st, 2nd, ..., $l$th channel probing of the relay. ${\mathbb E}[V^{\infty}(\lambda)]$ can be expressed as
\begin{multline*}
{\mathbb E}[V^{\infty}(\lambda)]=\text{Pr}[r_g^1\ge r_n](R_n\tau_d-\lambda \tau_2)
+\text{Pr}[r_g^1< r_n, r_g^2\ge r_n](R_n\tau_d-2\lambda \tau_2) + ...\\
+\text{Pr}[r_g^1<r_n,...,r_g^{l-1}<r_n,r_g^l\ge r_n](R_n\tau_d -l\lambda \tau_2)\\
+\text{Pr}[r_g^1<r_n,...,r_g^{l-1}<r_n,r_g^l< r_n] ({\mathbb E}[V^{\infty}(\lambda)]-l\lambda \tau_2).~~~~~~~~~~~~~~
\end{multline*}
So
\begin{multline}\label{e:compare_inf_l}
{\mathbb E}[V^{\infty}(\lambda)] - {\mathbb E}[V^l(\lambda)] = \text{Pr}[r_g^1<r_n,...,r_g^{l-1}<r_n,r_g^l< r_n]({\mathbb E}[V^{\infty}(\lambda)] -\lambda \tau_d) \\
=(F_g(r_n))^l ({\mathbb E}[V^{\infty}(\lambda)] -\lambda \tau_d) \ge 0~~~~~~~~~~~~~~~~~~~~~~~~~~~~~~~~~~~~
\end{multline}
which means the optimal second-hop strategy should be: the relay keeps probing the second-hop channel until the achievable rate is not less than $R_n$.

\subsection{Case with $\mathbb{E}[V^{\infty}(\lambda)] < \lambda\tau_d $}\label{s:vinfless}
If $\mathbb{E}[V^{\infty}(\lambda)] < \lambda\tau_d $, from (\ref{force_to_stop_or_not}) we have $\mathbb{E}[V^{\infty}(\lambda)] < \mathbb{E}[V^1(\lambda)]$. Now we compare $\mathbb{E}[V^1(\lambda)]$ with $\mathbb{E}[V^l(\lambda)]$, $l>1$.
\begin{equation*}
  \begin{split}
    &\mathbb{E}[V^{1}(\lambda)] - \mathbb{E}[V^{l}(\lambda)]\\
    =&-(\mathbb{E}[V^{\infty}(\lambda)]-\mathbb{E}[V^{1}(\lambda)]) + (\mathbb{E}[V^{\infty}(\lambda)]-\mathbb{E}[V^{l}(\lambda)])\\
    \overset{(a)}=& -\!F_g(r_n)(\mathbb{E}[V^{\infty}(\lambda)]\!-\! \lambda\tau_d)\!+\! (F_g(r_n))^l(\mathbb{E}[V^{\infty}(\lambda)]-\! \lambda\tau_d)\\
    =&F_g(r_n)\big(-1+(F_g(r_n))^{l-1}\big)(\mathbb{E}[V^{\infty}(\lambda)]- \lambda\tau_d) \overset{(b)}> 0
  \end{split}
\end{equation*}
in which $(a)$ comes from (\ref{force_to_stop_or_not}) and (\ref{e:compare_inf_l}), and $(b)$ comes from $F_g(r_n) < 1$ and $\mathbb{E}[V^{\infty}(\lambda)] < \lambda\tau_d $. Thus, the optimal second-hop strategy should be: the relay probes the second-hop channel only once, and transmits if the achievable transmission rate is not less than $R_n$, or gives up otherwise.

Overall, for the second hop, depending on comparison of $\mathbb{E}[V^{\infty}(\lambda)]$ with $\lambda \tau_d$, the relay should either probe the second-hop channel once, or keep probing the second-hop channel until the achievable second-hop rate is not less than $R_n$.

\section{Strategy for the First Hop}\label{SchemefortheFirstHop}
Based on optimal strategy in the second hop, now we derive optimal strategy for the first hop. In the first hop, at observation $n$, once the RTS of the winner source (i.e., the source that wins channel contention) is received by its relay, and the first-hop channel SNR denoted $r_f(n)$ is estimated, then the decision is either give-up or stop (i.e., to transmit), whichever has higher reward. If the decision for the first hop is give-up, then the net reward is $-\lambda \tau_{CTS}$ (since a CTS is needed to notify the decision); if the decision for the first hop is to transmit with rate $R_n$, the net reward is $ \max \, \{ \mathbb{E}[V^{1}(\lambda)],\mathbb{E}[V^{2}(\lambda)],\ldots,\mathbb{E}[V^{\infty}(\lambda)] \} - \lambda(\tau_{CTS} +\tau_d)$, in which $\tau_{CTS} +\tau_d$ is time cost in the first hop: the relay uses a CTS to notify the source of the decision and the source transmits with $\tau_d$ duration (noting that the time cost in the subsequent second hop is included in $\max \, \{ \mathbb{E}[V^{1}(\lambda)],\mathbb{E}[V^{2}(\lambda)],\ldots,\mathbb{E}[V^{\infty}(\lambda)] \}$).

First consider $\mathbb{E}[V^{\infty}(\lambda)] < \lambda\tau_d $ for the second hop.
Then based on discussion in Section \ref{s:vinfless}, $ \max \, \{ \mathbb{E}[V^{1}(\lambda)],\mathbb{E}[V^{2}(\lambda)],\ldots,\mathbb{E}[V^{\infty}(\lambda)] \} =  \mathbb{E}[V^{1}(\lambda)]$, so the net reward of transmission in first hop is
$   \mathbb{E}[V^{1}(\lambda)]- \lambda(\tau_{CTS} +\tau_d)$.
Since $\mathbb{E}[V^{\infty}(\lambda)] < \lambda\tau_d $, from (\ref{force_to_stop_or_not}) we have
\begin{equation}
\begin{array}{ll}
\mathbb{E}[V^{1}(\lambda)]& = (1-F_g(r_n))\mathbb{E}[V^{\infty}(\lambda)] + F_g(r_n) \lambda\tau_d\\
&<(1-F_g(r_n))\lambda\tau_d + F_g(r_n) \lambda\tau_d = \lambda\tau_d
\end{array}
\end{equation}
which leads to $\mathbb{E}[V^{1}(\lambda)]- \lambda(\tau_{CTS} +\tau_d)< -\lambda \tau_{CTS}$. In other words, the net reward of transmission in the first hop is less than the net reward of give-up in the first hop, and thus, the winner source will always give up in the first hop. Therefore, when we calculate the net reward of transmission in the first hop, we can ignore ``$\mathbb{E}[V^{\infty}(\lambda)] < \lambda\tau_d $". Thus, we focus on
$\mathbb{E}[V^{\infty}(\lambda)]\ge \lambda\tau_d$, and based on discussion in Section \ref{s:vinfmore}, we have $ \max \, \{ \mathbb{E}[V^{1}(\lambda)],\mathbb{E}[V^{2}(\lambda)],\ldots,\mathbb{E}[V^{\infty}(\lambda)] \} =  \mathbb{E}[V^{\infty}(\lambda)]$. So the net reward of transmission (stopping) in the first hop is
\begin{equation}\label{e:net-re}
  \begin{split}
    &\mathbb{E}[V^{\infty}(\lambda)]-\lambda(\tau_{CTS} +\tau_d)\\
    \overset{(c)}= & R_n\tau_d -\frac{1}{1-F_g(r_n)} \lambda\tau_2-\lambda(\tau_{CTS} +\tau_d) \\
    \overset{(d)}=  &\log_2(1+r_n)\tau_d - \lambda\tau_{CTS} - \lambda\tau_d - \lambda e^{\frac{r_n}{\rho_g}}\tau_2
  \end{split}
\end{equation}
in which $(c)$ comes from ${\mathbb E}[V^{\infty}(\lambda)]=R_n\tau_d -\frac{1}{1-F_g(r_n)} \lambda\tau_2$ which is from (\ref{e:sinfty_reward}), and $(d)$ is from
 $F_g(r_n)=1-e^{-\frac{r_n}{\rho_g}}$ (Rayleigh fading) and $r_n\triangleq 2^{R_n}-1$.
The net reward (\ref{e:net-re}) is not a monotonically increasing function of $r_n$. So we need to set up an optimal $r_n$ that makes the net reward maximal.

Define function $\phi(x) = \log_2(1+x)\tau_d - \lambda\tau_{CTS} - \lambda\tau_d - \lambda e^{\frac{x}{\rho_g}}\tau_2$, which is a concave function. To find the optimal $x$, denoted $x^*$, that maximizes $\phi(x)$, we can solve $\frac{d\phi(x)}{dx}=0$, which leads to
\begin{equation}
  \frac{\tau_d}{(1+x^*)\ln2} = \frac{\lambda}{\rho_g}e^{\frac{x^*}{\rho_g}}\tau_2.
  \label{r^*}
\end{equation}
$x^*$ can be calculated from (\ref{r^*}) numerically. So $r_n$ should be set to $x^*$ if feasible. However, it may not be feasible to set $r_n$ to be $x^*$ since $r_n$ should be no more than the first-hop channel SNR $r_f(n)$. Thus, overall we should set $r_n = \min\{r_f(n),x^*\}$ and $R_n = \log_2(1+\min\{r_f(n),x^*\})$.

Recall that an optimal stopping strategy of problem (\ref{e:transform}) with $\lambda^*$ satisfying $U(\lambda^*) = 0$ is an optimal stopping strategy of problem (\ref{e:original}). So next we focus on optimal stopping strategy of problem (\ref{e:transform}) with $\lambda^*$.
Maximal expected reward $U(\lambda^*)$ of problem (\ref{e:transform}) should satisfy an optimality equation \cite{ferguson2006optimal}:
\begin{multline*}
  \mathbb{E}\!\big[\max \big\{ \log_2\!(1+\min\{r_f(n),x^*\})\tau_d -\! \lambda^*(\tau_{CTS}+\tau_d + e^{ \frac{\min\{r_f(n),x^*\}}{\rho_g} }\tau_2),\\~
  U(\lambda^*)-\lambda^* \tau_{CTS} \big\}\big]- \lambda\tau_1 = U(\lambda^*).
\end{multline*}

Since $U(\lambda^*)=0$, the optimal equation is rewritten as
\begin{multline}
  \mathbb{E}\big[\max \big\{ \log_2(1+\min\{r_f(n),x^*\})\tau_d - \lambda^*(\tau_{CTS}+\tau_d + e^{ \frac{\min\{r_f(n),x^*\}}{\rho_g} }\tau_2),\\~
  -\lambda^* \tau_{CTS} \big\}\big]=\lambda^* \tau_1
  \label{optimalequation}
\end{multline}
from which $\lambda^*$ can be calculated numerically.

Accordingly, the optimal stopping strategy in the first hop is given as
{ \begin{multline}
  N^*(\lambda^*) = \min \big\{ n \geq 1:~  \log_2(1+\min\{r_f(n),x^*\})\tau_d - \lambda^*(\tau_{CTS}+\tau_d + e^{ \frac{\min\{r_f(n),x^*\}}{\rho_g} }\tau_2)\\ \ge
  -\lambda^* \tau_{CTS}\big\}
  \label{optimalstoppingrule}
\end{multline}}
in which $x^*$ can be calculated from (\ref{r^*}) with $\lambda=\lambda^*$. 

The left handside of the inequality in (\ref{optimalstoppingrule}) is a non-decreasing function of $r_f(n)$. Denote $\hat{r}_f$ as the solution of $r_f(n)$ for $\log_2(1+\min\{r_f(n),x^*\})\tau_d - \lambda^*(\tau_{CTS}+\tau_d + e^{ \frac{\min\{r_f(n),x^*\}}{\rho_g} }\tau_2)=
  -\lambda^* \tau_{CTS} $. Then the optimal stopping strategy in the first hop is rewritten as $N^*(\lambda^*) = \min \big\{ n \geq 1:~ r_f(n)\ge \hat{r}_f \big\}$. Thus, at observation $n$, if the fist-hop channel SNR $r_f(n)$ is less than the threshold $\hat{r}_f$, the winner source gives up; otherwise, the winner source stops, i.e., transmits with rate $R_n = \log_2(1+\min\{r_f(n),x^*\})$, and subsequently the relay keeps probing the second-hop channel until an achievable rate not less than $R_n$. The values of $\hat{r}_f$ and $x^*$ can be calculated offline, and thus, the optimal strategy is a pure-threshold strategy, with very low computational complexity.

\section{Performance Evaluation}\label{NumericalResult}
\begin{figure}[ht]
\begin{center}
\includegraphics[width=3.5in]{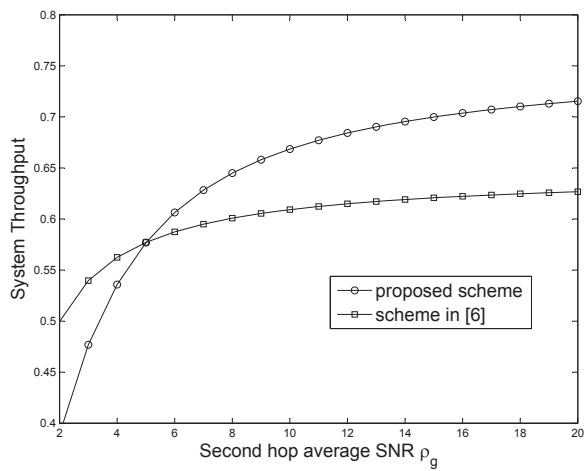}
\end{center}
\caption{System throughput of proposed scheme and the scheme in \cite{gong_xiaowen}.}
\label{comparison}
\end{figure}

Computer simulations are carried out to evaluate our proposed optimal relay-waiting scheme. The simulated network has 18 source-destination pairs, with other parameters set as: $\sigma = 20\text{$\mu$s}$, $\tau_{RTS} = 103 \text{$\mu$s}, \tau_{CTS}=\tau_{timeout} = 106\text{$\mu$s}$, $\tau_d = 0.8 \text{ms}$, $p = 0.1$, $\rho_f = 1$. We vary the average second-hop SNR $\rho_g$ from 2 to 20. For each $\rho_g$ value, we first numerically calculate $\hat{r}_f$, and use the value of $\hat{r}_f$ as the pure threshold in the simulations to obtain the system throughput. The simulation results are shown in Fig.~\ref{comparison}. Simulation results for the scheme in \cite{gong_xiaowen} (with direct links not considered) are also shown. It can be seen that, when the average second-hop SNR $\rho_g$ is below 5, the scheme in \cite{gong_xiaowen} achieves higher system throughput. But when $\rho_g>5$, our scheme achieves better throughput performance. Indeed, for a network with the second-hop channels having larger average SNR, if the current probed second-hop channel realization has a poor SNR, letting the relay wait may be more time-efficient compared to giving up.

\section{Conclusion}
In this letter, we exploited distributed opportunistic channel access in networks with DF relays, and we answered the question whether or not to let the relay wait for a better second-hop channel if the current second-hop channel is not good. For such relay-waiting, we derived the optimal strategies in the two hops. It was shown that the first-hop strategy is a pure-threshold strategy, while the second-hop strategy is to keep probing until a good enough channel is observed. Our simulations demonstrated that the optimal relay-waiting scheme should be adopted when the second-hop channels have larger average SNR.


\begin{thebibliography}{99}
\bibitem{zheng_dong}
D. Zheng, W. Ge, and J. Zhang, ``Distributed opportunistic scheduling
for ad hoc networks with random access: An optimal stopping approach,''
\emph{IEEE Transactions on Information Theory,} vol. 55, no. 1, pp. 205--222, Jan.
2009.

\bibitem{imperfect_information}
D. Zheng, M. O. Pun, W. Ge, J. Zhang, and H. V. Poor, ``Distributed
opportunistic scheduling for ad hoc communications with imperfect
channel information,'' \emph{IEEE Transactions on Wireless Communications,} vol. 7, no. 12,
part 2, pp. 5450--5460, Dec. 2008.

\bibitem{Ge_Zheng_2009}
W. Ge, J. Zhang, J.E. Wieselthier, and X. Shen, ``PHY-aware distributed
scheduling for ad hoc communications with physical interference model," {\it IEEE Transactions on Wireless Communications}, vol. 8, no. 5, pp. 2682--2693, May 2009.

\bibitem{Tan_Zheng_2010}
S.-S. Tan, D. Zheng, J. Zhang, and J. Zeidler, ``Distributed opportunistic scheduling for ad-hoc communications under delay constraints," in {\it Proc. IEEE INFOCOM 2010}.

\bibitem{two_level}
C. Thejaswi P. S., J. Zhang, M.-O. Pun, H. V. Poor, and D. Zheng, ``Distributed opportunistic scheduling with two-level probing,'' \emph{IEEE/ACM Transactions on Networking,} vol. 18, no. 5, pp. 1464--1477, Oct. 2010.



\bibitem{gong_xiaowen}
X. Gong, C. Thejaswi P. S., J. Zhang, and H. V. Poor, ``Opportunistic cooperative networking: To relay or not to relay,'' \emph{IEEE Journal on Selected Areas in Communications,}
vol. 30, no. 2, pp. 307--314, Feb. 2012.

\bibitem{zhangzhou}
Z. Zhang and H. Jiang, ``
Distributed opportunistic channel access in wireless relay networks,'' \emph{IEEE Journal on Selected Areas in Communications,}
vol. 30, no. 9, pp. 1675--1683, Oct. 2012.

\bibitem{ferguson2006optimal}
T.~Ferguson, {\it Optimal stopping and applications}, Available online: {http://www.math.ucla.edu/{\texttildelow}tom/Stopping/Contents.html}, Mathematics Department, UCLA.

\end{thebibliography}
\end{document}